\documentclass[jbr]{ipart}

\RequirePackage{hyperref}

\usepackage[utf8]{inputenc} 
\usepackage[T1]{fontenc}    
\usepackage{hyperref}       
\usepackage{url}            
\usepackage{booktabs}       
\usepackage{amsfonts}       
\usepackage{nicefrac}       
\usepackage{microtype}      
\usepackage{natbib}

\usepackage{standalone}

\usepackage{graphicx}
\usepackage{float}
\usepackage{multirow}
\usepackage{geometry}
\usepackage{amsmath}
\usepackage{amssymb}
\usepackage{siunitx}
\usepackage{diagbox}
\usepackage[shortlabels]{enumitem}

\usepackage[dvipsnames]{xcolor}
\usepackage{multirow}

\usepackage{tikz}
\usetikzlibrary{positioning}

\newcommand{\argmin}{\mathop{\mathrm{argmin}}}          
\newcommand{\argmax}{\mathop{\mathrm{argmax}}}          

\usepackage[ruled,vlined]{algorithm2e}

\usepackage{subcaption}
\usepackage{pifont}
\startlocaldefs
\theoremstyle{plain}

\endlocaldefs

\pubyear{2026}
\volume{0}
\issue{0}
\firstpage{1}
\lastpage{13}
\arxiv{2603.16949}

\begin{document}

\begin{frontmatter}
\title[Entropy-Aware Task Offloading in MEC]{Entropy-Aware Task Offloading in Mobile Edge Computing}

\begin{aug}
    \author{\inits{M.}\fnms{Mohsen} \snm{Sahraei Ardakani}\thanksref{t2}\ead[label=e1]{msahrae@ncsu.edu}},
    \address{Dept. of Electrical and Computer Engineering\\
    North Carolina State University\\
             Raleigh, NC 27606\\
             \printead{e1}}
    \author{\inits{R.}\fnms{Rui} \snm{Song}\ead[label=e2]{rsong@ncsu.edu}}
    \address{Dept. of Statistics\\
North Carolina State University\\
Raleigh, NC 27606\\
             \printead{e2}}
    \and
    \author{\inits{H.}\fnms{Hong} \snm{Wan}
            \ead[label=e3]{hwan4@ncsu.edu}%
           }
    \address{Dept. of Industrial and System Engineering\\
North Carolina State University\\
Raleigh, NC 27606\\
             \printead{e3}}
    \thankstext{t2}{Corresponding author.}
\end{aug}
%

\begin{abstract}
Mobile Edge Computing (MEC) technology has been introduced to enable could computing at the edge of the network in order to help resource limited mobile devices with time sensitive data processing tasks. In this paradigm, mobile devices can offload their computationally heavy tasks to more efficient nearby MEC servers via wireless communication. Consequently, the main focus of researches on the subject has been on development of efficient offloading schemes, leaving the privacy of mobile user out. While the Blockchain technology is used as the trust mechanism for secured sharing of the data, the privacy issues induced from wireless communication, namely, usage pattern  and location  privacy are the centerpiece of this work. The effects of these privacy concerns on the task offloading Markov Decision Process (MDP) is addressed and the MDP is solved using a Deep Recurrent Q-Netwrok (DRQN). The Numerical simulations are presented to show the effectiveness of the proposed method.
\end{abstract}



\end{frontmatter}


\maketitle

\section{Introduction}

The Mobile Edge Computing (MEC) provides the cloud computing services at the edge mobile user network. MEC has emerged as a solution for distributed processing of time sensitive tasks  closer to the cellular users with a wide range of applications including  healthcare, Traffic Management, and Gaming to name a few\cite{hu2015mobile, mach2017mobile, ahmed2017mobile, feng2017ave}.
The limited computation power due to strict power and size constraints in situations similar to health care Internet of Things (IoT)  make it vital rely on cloud computing resources for their data processing demand, nevertheless due to time sensitivity of these data processing tasks the conventional cloud computing concepts would not be a viable option. The MEC technology delivers faster processing time by exploiting  idle computation resources at the edge of the network  therefore provides the computational resources and at the same time satisfies the processing time constraints\cite{islam2015internet, abbas2017mobile, chen2015efficient}. However extending the cloud computing at the edge of the network  requires establishment of trust between the mobile user and the MEC servers. Blockchain empowered MEC addresses the issue of trust between the edge devices\cite{xu2019become}. The distributed computing network is consisted of three layers namely, mobile devices, edge servers and cloud servers. While the edge servers provide mobile users with their  processing power demand, the cloud servers guarantee the timely mining of blocks and blockchain formation\cite{luo2020blockchain}. Unlike the conventional cloud computing where a third party manages the permissions to access and decrypt the data, in blockchain empowered MEC  the access controls  are kept in the blockchain. The blockchain technology gurantees the anonymity by design\cite{xiong2018mobile}. Additionally,  since both task assignment and completion timestamps are stored on the blockchain, edge server computation power under-delivery will be punished and a certain level for Quality of Service will be guaranteed\cite{wu2020cooperative}, \cite{xiao2020reinforcement}. On the other hand the adoption of blockchain technology in the distributed computing system requires efficient resource allocation for handling of the mining tasks, data processing tasks and cloud  service purchases at the edge servers layer in order to keep their participation profitable\cite{liu2010tradeoff, xiong2018cloud}.

Similarly, the IoT device needs to decide how to assign processing  the edge and local resources in order to achieve longer battery life and lower delay. This problem is known as optimal offloading policy. The mobile user offloading policy determines the partition of task that are going to be offloaded to the MEC servers or processed locally. 

There has been ongoing research on the solution of optimal task offloading decision which minimizes the energy costs and processing time. The wireless communication  turns to be the most decisive factor in the optimal decision\cite{mao2017survey, you2016multiuser}.It has been shown that in non private optimal policies the IoT device tends to offload all of the tasks to the closest server in good channel condition and processes everything locally in bad channel quality\cite{ dai2018learning}.
Additionally,  due to limited computation power at the IoT devices it is not practical to use elaborate encryption methods in  the wireless communications, therefore, task offloading becomes prone to information leaks and consequently compromised privacy\cite{mach2017mobile}.  Given the potential of MEC technology and fog computing there has been a number of researches aiming at addressing the issue of privacy in IoT, moreover there has been ongoing discussion on the design element that must be built in the MEC network and communications within the network in order to avoid certain risks\cite{porambage2016quest, ren2018querying}. A wide range of privacy aspects of a device in IoT namely, identity, data, location, are reviewed in \cite{alrawais2017fog}. Besides these issues there are certain privacy breaches such as authentications, data sharing permissions, ... due to the wireless communications in the MEC network\cite{yi2015security}. The concept of privacy for the mobile user has a broader meaning. The mobile user prefers to interact with secure and computationally reliable servers based on this idea  a concept of server trustworthiness is introduced in \cite{wu2019trust}, where the user learns to find a subset of reliable servers among available servers with these criteria. On the other side the nodes that act as server have a similar problem. The servers need to dynamically determine the price of computational resources  in order to both avoid overflow at server and maximize their profit. The servers need to learn a pricing policy that does not depend on the user's sensitive data in order preserve their privacy, otherwise the user will drop to connections to the server\cite{li2019learning}. 

A number of researches have specifically explored the privacy issues with regards to  the wireless communication in MEC, however, they have failed to make a mathematical case for privacy evaluation\cite{he2019physical,zhao2020privacy}.  In \cite{he2017privacy}, a heuristic privacy metric is introduced to address the information leaks at a compromised server.  The idea behind this privacy metric  is to add an incentives for non optimal decisions and get a less predictable action as a result. Then, the metric is used to formulate the task offloading decision as a Constrained Markov Decision Process (CMDP). This approach  leads to a one step optimization problem which fails to capture the long term patterns. The information leaks and privacy issues in wireless communication with a compromised server have been subject of a number of researches. Given that the distance between the server and the user highly correlates with wireless channel gain, and the data mobile user's data usage can be used as a fingerprint to identify the mobile user, the adversary is enabled to identify a user and estimate their location by compromising a MEC server\cite{he2018deep}. Altogether, an agent needs to learn the   optimal task offloading decision under the privacy constraints. 

Most recent works on the optimal task offloading problem, model the MEC network as  a Markov Decision Process(MDP) and learn the optimal offloading policy  with  a Reinforcement Learning(RL) method\cite{luo2020collaborative}. Application of RL methods are limited by the size of the action and state spaces. As the size of state and action spaces increases the Q-learning\cite{sutton2018reinforcement} fails to converge to the optimal solution and Deep Q-Network(DQN) is proved to be more effective\cite{nguyen2020privacy}. To overcome the action space size a genetic algorithm optimization is employed in which resulted in faster convergence\cite{qiu2019online}. 

In this work the privacy reward is formulated based on the long term decisions of the IoT device. It is claimed that by aiming towards independence of leaked information and hidden  variables, usage pattern and location could be kept private. It has been shown that the entropy reward function is reduces the probability of accurate adversary attack. Furthermore, given the non-markovian nature of the entropy function a model based on Deep Recurrent Q-Network(DRQN) is proposed to solve the reward function. The numerical results show the DRQN model  outperforms the previously used DQN models and successfully learns the entropy reward function.\\
The main contributions of this work include, introducing a novel privacy reward function, making a mathematical case for effectiveness of the privacy function in the MEC network, and proposing a DRQN  model for solving the non-markovian privacy reward.
The remainder of this paper is organized as follows. In Section \ref{sys} the MEC network is modeled as an MDP and task offloading problem with privacy concern is defined. In Section \ref{sec:sol}, the reinforcement learning algorithm used is explained. In Section \ref{sec:sim}, the simulated results are presented and discussed. Section \ref{sec:conc} concludes this paper with future direction and potential applications.

\section{System Model and Problem Formulation}\label{sys}
\subsection{System Model}
    The MEC technology enables the IoT devices to use the processing power of idle devices in the network acting as MEC servers   leading to longer battery life and reduced  latency. In order to formulate the MEC network it is assumed that the time is slotted at equal intervals and at each time slot a number of data processing tasks is generated by the IoT device. The IoT device can process the tasks locally, offload them to the server or put them in the buffer for future time slots. The objective of the IoT device is to use an offloading policy that  achieves energy efficiency, timely processing and at the same time does not jeopardise its privacy.
For a single IoT device that is connected to a single MEC server, the task offloading decision is modeled as an MDP in order to facilitate the use of RL algorithms. MDP is defined as a tuple $(\mathcal{S, A, P, R})$ where $\mathcal{S}$, $\mathcal{A}$ denote state and action spaces, $\mathcal{P}(s^\prime| s , a)$ determines the transition probability of going to state $s'\in \mathcal{S}$ given the current state $s\in \mathcal{S}$ and action taken $a\in \mathcal{A}$.
$\mathcal{R}(s,a)$ represents the immediate reward  from taking an action $a\in \mathcal{A}$ at the state $s\in \mathcal{S}$.

\begin{table}[t!]
  \caption{List of notations}
\small
    \centering
 \begin{tabular}{ p{.05\textwidth} p{.37\textwidth}}
 \toprule
\centering \scriptsize Symbol& Description\\\hline
 $d_n$ & number of new tasks at timestep n\\
 $b_n$ & number of  tasks in the buffer at timestep n\\
 $g_n$ & channel quality at timestep n\\
 $q_n$ & number of buffered tasks  at timestep n\\
 $t_n$ & number of offloaded tasks at timestep n\\
 $l_n$&number of tasks processed locally at timestep n\\
 $r$& transmission speed\\
 $e_0(e_1)$& offloading energy per task in bad(good) channel condition\\
 $M$& task size\\
 $F$& local CPU frequency\\
$\eta$& task workload density\\
$E$& Energy consumption\\
$L$& task processing time\\
$P$& privacy reward\\
\bottomrule
\end{tabular}
    \label{tab:notation}
\end{table}

\subsection{MDP Formulation}
Based on the assumption of time being slotted, we describe the details of the MDP formulation of the MEC network. The MDP formulation of the MEC network is described in the rest of this section. The state and action spaces and the transition probabilities  closely follows the MEC MDP model devised in \cite{he2018deep}.   \\ \indent
  The state variable $s_n = (d_n, b_n, g_n) \in \mathcal{S}$  at timeslot $n$ is defined as a tuple of wireless channel gain of $g_n \in \{0, 1\}$,    number of data processing tasks in the buffer of the device ($b_n \in \{0, 1, \dots  b_\text{max}\}$) and a number of newly  generated data processing tasks ($d_n \in \{0, 1, \dots d_\text{max}\}$).\\ \indent  The device has to decide to how to distribute those tasks over the available computation resources, thus the action variable $a_n = (q_n, t_n) \in \mathcal{A}$ determines how many tasks are going to be offloaded to the server ($t_n \in \{0, 1, \dots d_\text{max} + b_\text{max}\}$) and stored in the buffer for future ($q \in \{0, 1, \dots  b_\text{max}\}$) time slots. Consequently, the rest will be processed locally ($l_n = b_n+d_n - q_n - t_n$).
The immediate reward $\mathcal{R}: \mathcal{S}\times \mathcal{A} \to \mathbb{R}$ is defined such way that reflects the ultimate goals of the IoT device which are low energy consumption and near real time processing. Furthermore, the reward function can incorporate other  objectives like privacy. The agent will learn to maximize the total collected reward over a long period of time. Thus, achieving those goals. Contrary to the energy consumption and processing the privacy is not well defined. In our work the  entropic privacy reward is introduced for the first time and proved its efficacy in measuring risk of certain adversary attacks.

 As mentioned previously, the IoT device has three options to handle the data processing tasks. If the IoT device offloads $t_n$ tasks  the completion time with regards to downloading the results and server processing time is negligible. Therefore the latency only depends on the transmission time shown in Equation~\ref{eq:off_latency}. Moreover, since  the transmission speed $r$ is supposed to be fixed the IoT device needs  to compensate for the channel quality. Therefore, according to the channel condition in good (bad) channel quality the the transmission energy per task is specified as $e_1$ ($e_0$). The Equation~\ref{eq:off_energy} show the offloading energy consumption\cite{you2016energy}.

\begin{align}
    L^\text{offload} &=t_n\frac{M}{r}  \label{eq:off_latency}\\
    E^\text{offload} &= e_{g_n}t_n  \label{eq:off_energy}
\end{align}
On the other hand for local processing decision of  $l_n$ tasks  at time t, the completion time is determined based on the task workload density $\eta$ ($cycle/kb$), task size $M$ ($kb$) and local computation power F ($cycle/sec$) given in Equation~\ref{eq:loc_latency}. The local processing energy is a function of  CPU frequency and efficiency which can be further reduced to a constant $e_l$ for  local energy consumption per task as Equation~\ref{eq:loc_energy}.
\begin{align}
    L^\text{local} &=l_n\frac{M\eta}{F}  \label{eq:loc_latency}\\
    E^\text{local}&= e_ll_n  \label{eq:loc_energy}
\end{align}
The last option for the IoT device is to put $q_n$ tasks in the buffer which the does not need any energy, however it adds the duration of one time slot ($\Delta t$) to the latency computed for that timestep.
Since the local processing and edge processing work in parallel the total latency is the maximum of local processing and offloading latencies add to the queuing delay.  Total energy is computed as the sum of its decision level macros, shown in Equation~\ref{eq:latency}-\ref{eq:energy}\cite{he2017privacy}.
\begin{align}
    L(a_n) &= q_n\Delta t + \max(L^\text{offload},L^\text{local} ) \label{eq:latency}\\
    E(s_n, a_n) &= E^\text{offload} + E^\text{local}  \label{eq:energy}
\end{align}
Since  the transmission speed $r$ is supposed to be fixed the IoT device needs  to compensate for the channel quality. Therefore, in good (bad) channel quality the variable $e_1$ ($e_0$) is defined as the transmission energy per task.
Consequently, the immediate reward collected from taking an action   is defined as the difference between the privacy rewards (yet to be discussed) and action costs. Since it is not possible to necessarily minimize both energy consumption and latency the constant $w_q$ balances the trade-off between those two objectives in Equation~\ref{eq:cost}\cite{he2017privacy}.

\begin{align}
    C(s_n, a_n) &= w_q L(a_n) + E(s_n, a_n) \label{eq:cost}\\
    r(s_n, a_n) &= \lambda P(s_n, a_n) - C(s_n, a_n) \label{eq:reward}
\end{align}\\
Hence, the objective of the IoT device is to adopt the decision policy $\pi$ that maximizes its average long-time average expected reward given in Equation~\ref{eq:expected}.
\begin{align}
    \bar R &= \lim_{N\to \infty} \frac{1}{N}\mathbb{E}_\pi [\sum_{\tau=0}^N r(s_\tau, a_\tau)] \label{eq:expected}
\end{align}

\begin{figure*}[t!]
\begin{subfigure}[]{.27\textwidth}
    \centering
    \includegraphics[ scale =1 ,angle =90]{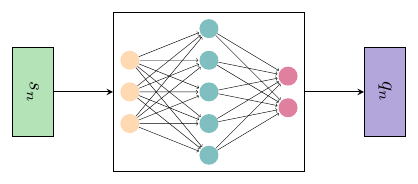}
    \caption{DQN}
    \label{fig:dqn}
\end{subfigure}
\begin{subfigure}[]{.67\textwidth}
    \centering
    \includegraphics[  scale =1,angle = 90]{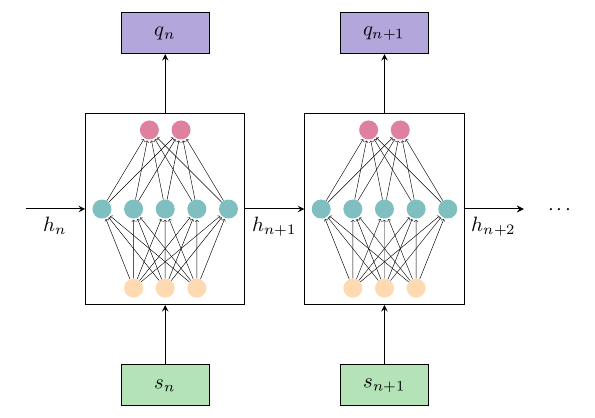}
    \caption{DRQN}
    \label{fig:drqn}
\end{subfigure}\caption{Comparison of neural network models used in DQN and DRQN}\label{fig:DQN1}
\end{figure*}

\subsection{Privacy Concern}
There has been a number of researches aiming at finding the optimal offloading policy under different assumptions\cite{qiu2019online,luo2020collaborative,nguyen2020privacy}. However, the privacy issue in MEC networks has not been fully explored and there has been ongoing efforts to make wireless communication more secure. The   tasks offloaded to the server is at risk of being observed by the adversary in the wireless communication between the IoT device and the server\cite{dai2018learning}. It has been shown that in non private optimal policies the IoT device tends to offload all of the tasks to the server in good channel condition and processes everything locally in bad channel quality\cite{he2017privacy}. Additionally, the channel condition is highly correlated with the distance between the IoT device and the Edge server\cite{nguyen2020privacy}.  This paves the way for the adversary to detect and estimate the location of a targeted IoT device by compromising the server and observing the offloading decisions. Therefore both data usage pattern and location of the IoT device are sensitive information that could be leaked in the MEC network. Data usage pattern can be used as a fingerprint to identify the IoT device. The adversary observes the offloading behaviour of a certain IoT device at the server and compares it against the recorded usage pattern of the target user and if a match were detected the location would be estimated based on the location of the server\cite{he2018deep}.
Intuitively, in order for the IoT device to preserve its privacy it is necessary to avoid the apparent cost effective decision and make the data usage pattern less predictive of the actual data generation distribution. Therefore, adding a privacy term in the reward function will push the agent towards a policy that incorporates that privacy metric.  Since the mechanism used by the adversary for inferring sensitive information is widely unknown, the choice for privacy reward function has remained heuristic\cite{he2017privacy}. However, there is the potential to make a clear case for the privacy reward function under certain assumptions. In the case of a compromised edge server the adversary can observe the variable $o_n$ and the user needs to keep their sensitive information such as $d_n$ and $g_n$ secure. Ultimately, as mentioned before the channel condition estimation helps the adversary with the target location estimation, and $p(d)$ enables target detection.\\
\indent \quad\textit{Remark 1:} The offloading policy  maximizing $H(D,T)$ minimizes the probability of correct $d$ estimation for the adversary.\\
\textit{proof:} If the adversary has the estimation of $\hat d$ we have:

\begin{align}
    p(\hat d =d) &=  \sum_{t} p(\hat d =d|t)p(t)  \nonumber\\ &\leq \sum_{t}\max_{d'}(p(d'|t))p(t)\nonumber\\
    & =\sum_{d,t}\max_{d'}(p(d'|t))p(d,t)  \nonumber \\
     &= E[\max_{d'}(p(d'|t))]
\end{align}\\
\noindent The agent must learn a policy that lowers the upper bound for correct estimation probability, therefore:
\begin{multline}
    \argmin_{p(d|t)} E[\max_{d'}(p(d'|t))]\\ =\argmin_{p(d|t)}
    \log (E[\max_{d'}(p(d'|t))] )
\end{multline}
\begin{align}
    \log (E[\max_{d'}(p(d'|t))] ) &\geq  E[\log (\max_{d'}(p(d'|t)))] \nonumber\\&\geq E[p(d|t)]\nonumber\\ &= - H(D|T)\\
    \Rightarrow \argmin_{p(d|t)} E[\max_{d'}(p(d'|t))]& =\argmax_{p(d|t)} H(D|T)
\end{align}

 The same argument holds for the $g$ estimation. Additionally since the estimated marginal distribution $T$ could be used for identification purposes \cite{he2019peace}, if the user maximizes the variation in the offloading decision it would further complicate the adversary attack.   Therefore the privacy reward in Equation~\ref{eq:reward} has to be the sum of theses  terms, resulting in the privacy reward of Equation~\ref{eq:privacy}. According to the proof for any adversary attack the privacy reward in Equation~\ref{eq:privacy} will keep $d$ and $g$ secure.
\begin{align}
    P(s, a) &= H(D|T) + H(G|T) + H(T) \nonumber\\
    &= H(D,G,T) \label{eq:privacy}
\end{align}
It should be noted that since the joint distributions are not known, both the distributions and consequently the entropies are approximated using the histogram of past W events. Thus, the joint probability of a tuple  $(d, g, t)$ at time step $n$ with $m$ occurrences in the previous $W$ state and action pairs,$\{(s_{n-W+1}, a_{n-W+1}),..., (s_{n}, a_{n})\}$,  is approximated as its frequency according to Equation~\eqref{eq:hist}.
\begin{align}
    \hat p_n(d, g, t) &= \frac{m}{W} \label{eq:hist}
\end{align}

The efficacy of entropy reward function and its approximation was discussed in this section, however, since the entropy reward function is approximated using past states and actions it could be argued that the reward is no longer markovian and the MEC model is no longer an MDP. Given that, the entropy reward function is  a function of past $W$ states and actions, the state variable is  changed according to Equation~\eqref{eq:state_new} to keep the MEC model as an MDP. This modification will resolve the issue of non markovian  reward. However, it will make  the size of the state space $|\mathcal{S}'| = |\mathcal{S}|^W \times |\mathcal{A}|^{W-1}$, to grow exponentially with $W$, thus it would not be feasible for the agent to learn the entire state space. Furthermore the state is highly redundant. Therefore under a Partially observable Markov Decision Process(POMDP) that the agent only observes current state as previously defined the size of the state space would not be a challenge as long as the agent can learn temporal patterns. A POMDP is a 7-tuple of   $(\mathcal{S', A, P, R}, \Omega, \mathcal{O})$ where in addition to MDP components $\Omega$ represents the finite set of observations that the mobile user can experience which will be $\mathcal{S}$ as defined previously and $\mathcal{O}: \mathcal{S'} \to \Pi(\Omega)$ is the observation function that gives the probability of making an observation $o$ in state $s$\cite{kaelbling1998planning}. However, in the case of MEC problem with entropy reward it will be deterministic and the observation will be the $s_{n}$ as defined in Equation~\eqref{eq:obs_fun}.
\begin{align}
     s'_n &= (s_{n-W+1}, a_{n-W+1}, ..., s_n) \in \mathcal{S}' \label{eq:state_new}
\end{align}
\begin{align}
     \mathcal{O}(s_n', o_n) &= \mathcal{O}((s_{n-W+1}, a_{n-W+1}, ..., s_n), o_n)\nonumber\\ &= \begin{cases}
     1 \mbox{ if } o_n = s_n\\
     0 \mbox{ otherwise }
     \end{cases}\label{eq:obs_fun}
\end{align}
The transition probabilities and action space will remain the same and the entropy reward function in Equation~\eqref{eq:privacy} will be approximated using Equation~\eqref{eq:hist} and the  immediate reward function will be determined according to Equation~\eqref{eq:reward}.

\section{Proposed Solution} \label{sec:sol}
The main contribution of this work is to improve the performance of the MEC from the aspect of privacy concerns. In the previous section, it is shown that the entropic privacy reward creates the means necessary to evaluate  the privacy of the user in the MEC network. Consequently, an agent trained on that reward function will behave more privately. The idea behind  Deep Q-Network (DQN) algorithm is to use a neural network to learn the  Q-function $Q:|\mathcal{S}| \times |\mathcal{A}|\to \mathbb{R}$\cite{mnih2015human} that is the expected discounted reward using a policy $\pi$ starting from state $s$ and taking action $a$ as defined in Equation~\eqref{eq:qlearning1}\cite{sutton2018reinforcement}.
\begin{equation}Q^\pi(s, a)  = E_\pi[\sum_{k=0} \gamma^k R_{n+k+1} |s_n = s, a_n = a]\label{eq:qlearning1}
\end{equation}

\begin{algorithm}[t]
\SetAlgoLined
\SetKwInOut{Input}{input}
\SetKwInOut{Output}{output}
\Input{ Neural Network model, discount factor $\gamma$, positive integer $\#episodes$, learning rate $\alpha$ and $\epsilon_i$ sequence, $\tau$ polyak averaging weight}
\Output{trained model parameters $\phi$ which determines Value function $Q_{\phi}$ ($\approx q_\pi$)}

 initialize $\phi$ arbitrarily\;
 $\phi'\leftarrow \phi$\;
 \For{$i \leftarrow 1$ to \#episodes}{
  $\epsilon \leftarrow \epsilon_i$\;
 Collect dataset $\{(s_i, a_i, s'_i, r_i)\}$ using policy derived from $Q_\phi$ ($\epsilon-greedy$) add it to $\mathcal{B}$\;
 Sample a batch from $(s_i, a_i, s'_i, r_i)$ from $\mathcal{B}$\;
 $\phi \leftarrow \phi -\alpha \sum_i \frac{d\, Q_\phi}{d\, \phi}(s_i, a_i)[Q_\phi(s_i, a_i) - (r(s_i, a_i)+ \gamma \max_{a'} Q_{\phi'}(s'_i, a'_i))]$\;
 $\phi' \leftarrow \tau \phi' +(1- \tau)\phi$\;
 }
 \Return {$\phi$}
 \caption{DQN algorithm}\label{alg:dqn}
\end{algorithm}

 The set of neural network parameters $\phi'$ are learnt through the backpropagation algorithm by minimizing the loss between the  target values and neural network output. On the other hand in the MDP  the  target values for neural network are not explicitly given, hence the target values are imitated using another structurally identical neural network. There are three steps in the DQN algorithm:
\begin{itemize} \item Data collection and eviction
\item Batch sampling and Backpropagation ($\phi$ update)
\item Target neural network update ($\phi'$ update)
\end{itemize}
Throughout the training, the agent interact with the environment according to the $\epsilon - greedy$ policy built on $Q_\phi$, and based on those interactions the policy network and target network are updated according to Algorithm~\ref{alg:dqn}.
\begin{figure*}[t]
\includegraphics[trim={0 0 0 1.2cm},clip,width = 1\textwidth]{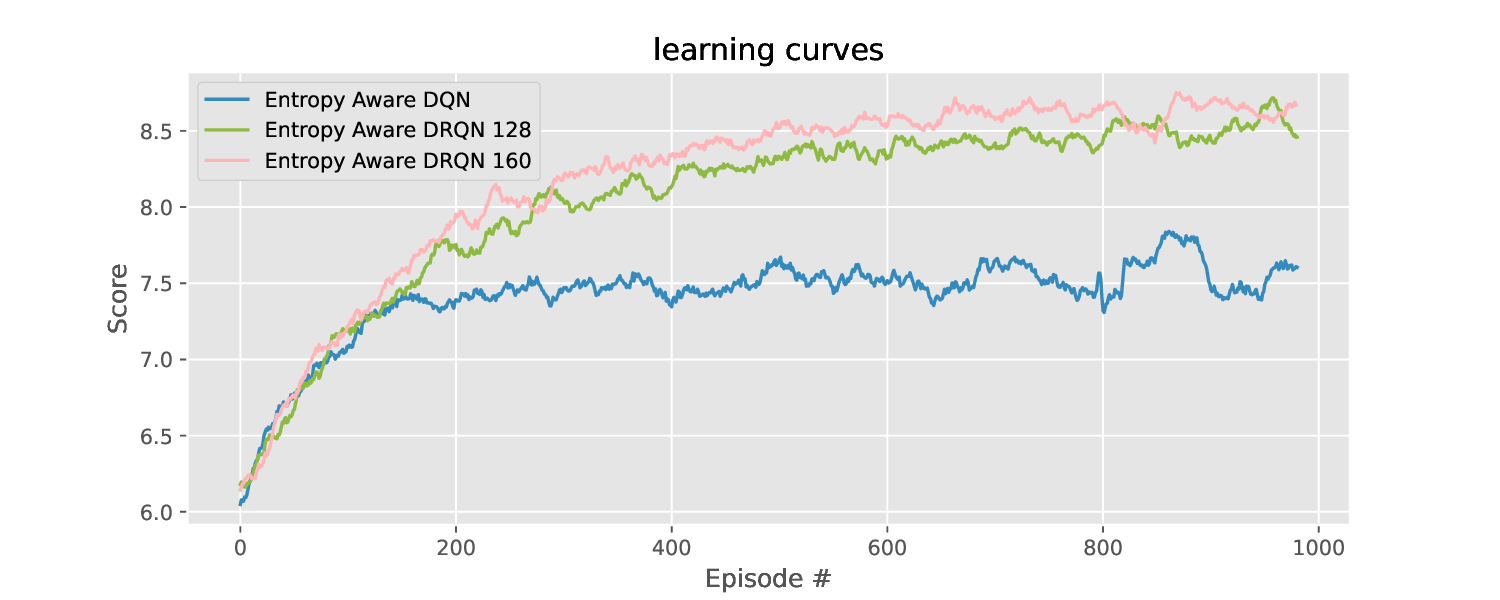}
     \caption{Learning Curves of the trained agents}
    \label{fig:lc}
\end{figure*}

\noindent Given the modifications in the MDP,  size of the state space $|\mathcal{S}'| = |\mathcal{S}|^W \times |\mathcal{A}|^{W-1}$,  exponentially grows with $W$ and make it infeasible to train an agent using  Deep Q-Network(DQN) algorithm. In order to reduce a the size state space it is assumed that the state space is partially observable, and the agent only observes the current state ($s_n$). As a result, it is necessary for the agent to have some memory of the past to be able to learn the reward function and act accordingly. Therefore, the neural network in DQN  specified in Algorithm~\ref{alg:dqn} is replaced with a Recurrent Neural Network (RNN) resulting in Deep Recurrent Q-Network (DRQN) algorithm\cite{hausknecht2015deep}. As shown in Figure~\ref{fig:DQN1}, the RNN used in DRQN outputs some feedback to the network for the next time step in addition to the offloading decision. Therefore, the network has the potential to learn temporal patterns hence reconstruct the entropy reward function. There is one consideration that contrary to the DQN algorithm the training samples are not the tuples of state, action, next state and immediate reward but a trajectory of a states and actions and the immediate reward for the last action.

\begin{table}[t]
  \caption{DQRN parameters}
\small
    \centering
 \begin{tabular}{ p{.25\textwidth} p{.18\textwidth}}
 \toprule
\centering Parameter& Value\\\hline
 Learning episodes & $10^3$\\
 Discount Factor  $\gamma$ & $0.9$\\
 Exploration rate $\epsilon$ & $1\text{ with } 0.995$ decay\\
 Recurrent Q-Network layers&$6$\\
 {RNN layer Type (hidden units)} &{3$\times$GRU(128), 2$\times$FC(128),  FC($|\mathcal{A}|$)}\\
 Replay buffer size& $10^5$\\
 Batch size & $128$\\
 Soft update factor (period target) $\tau$ & $0.0001$ (2)\\
Training Sequence Length & 128(160)\\
\bottomrule
\end{tabular}
    \label{tab:sys_param}
\end{table}
\section{Simulation Results}\label{sec:sim}

\begin{figure*}[t]
\begin{subfigure}[]{.32\textwidth}
    \centering
    \includegraphics[trim={0 0 0 1.15cm},clip,width = 1\textwidth]{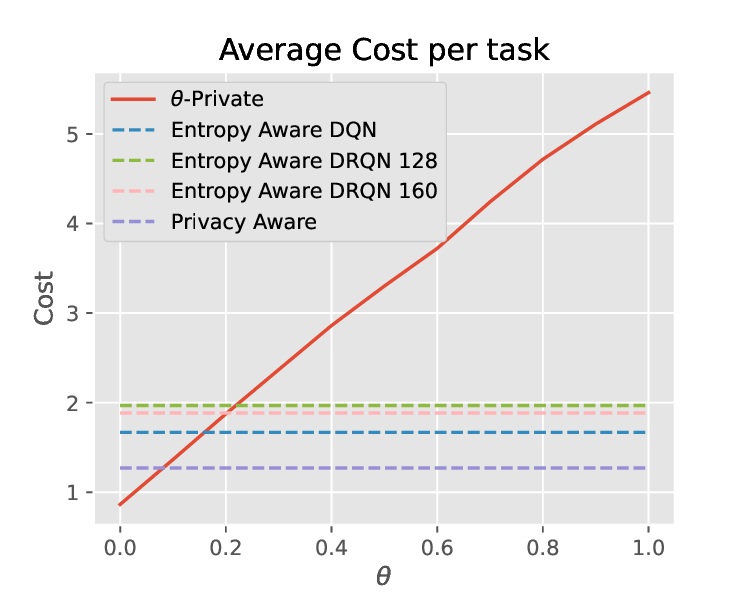}
    \caption{cost}
    \label{fig:cost}
\end{subfigure}
\begin{subfigure}[]{.32\textwidth}
    \centering
    \includegraphics[trim={0 0 0 1.15cm},clip,width = 1\textwidth]{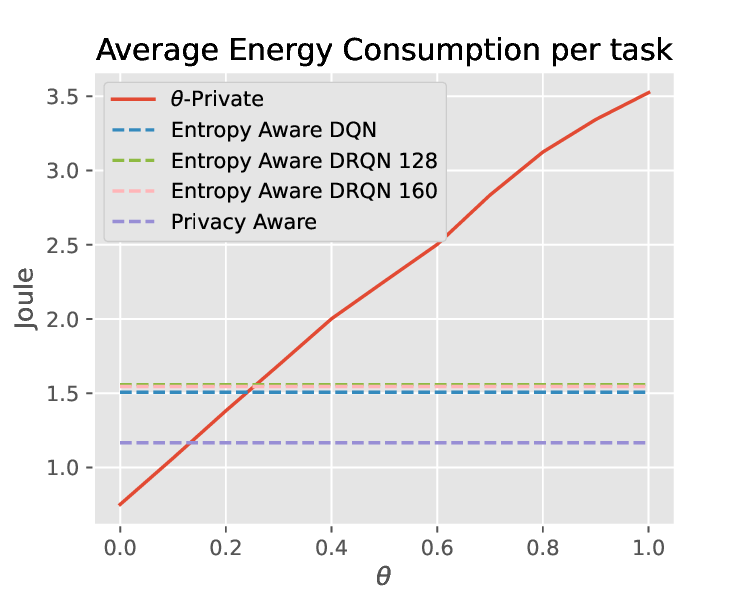}
    \caption{Energy Consumption}
    \label{fig:energy}
\end{subfigure}
\begin{subfigure}[]{.32\textwidth}
    \centering
    \includegraphics[trim={0 0 0 1.15cm},clip,width = 1\textwidth]{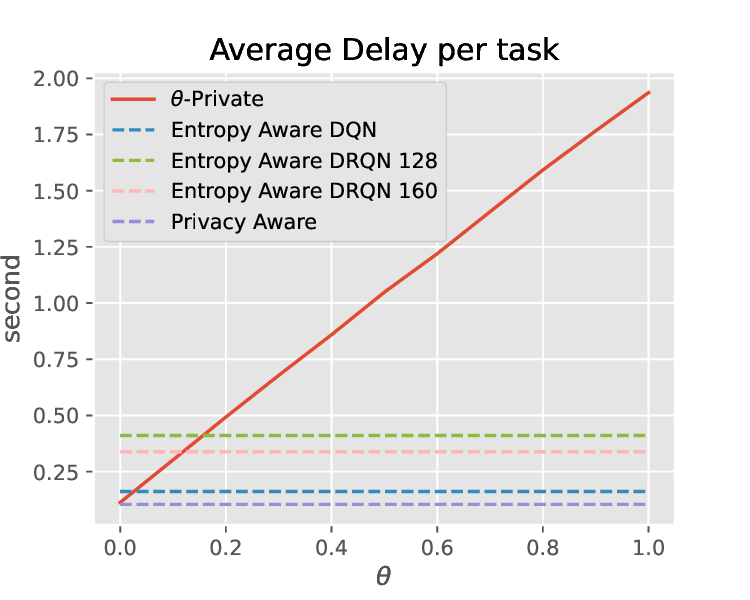}
    \caption{Delay}
    \label{fig:delay}
\end{subfigure}\caption{Comparison of reward macros for the agents}\label{fig:res1}
\end{figure*}
\begin{figure*}[t]
\begin{subfigure}[]{.32\textwidth}
    \centering
    \includegraphics[trim={0 0 0 1.15cm},clip,width = 1\textwidth]{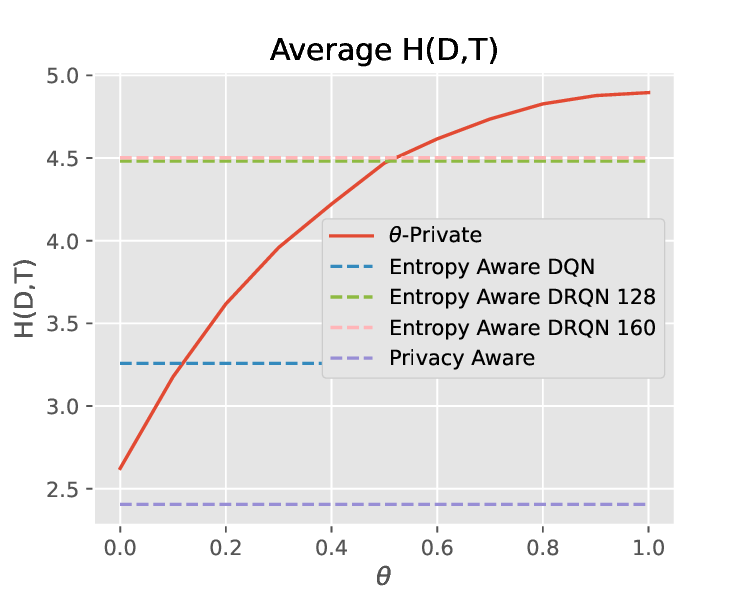}
    \caption{Joint data usage entropy}
    \label{fig:Hdt}
\end{subfigure}
\begin{subfigure}[]{.32\textwidth}
    \centering
    \includegraphics[trim={0 0 0 1.15cm},clip,width = 1\textwidth]{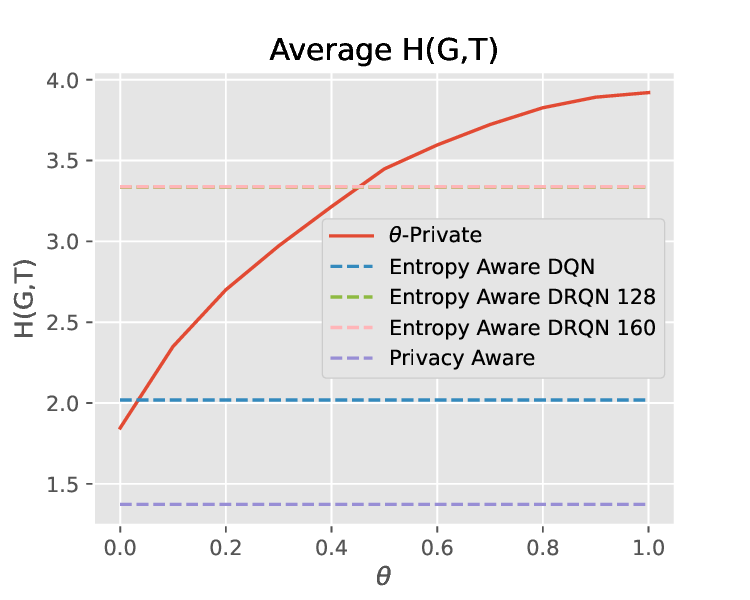}
    \caption{Joint channel entropy}
    \label{fig:Hgt}
\end{subfigure}
\begin{subfigure}[]{.32\textwidth}
    \centering
    \includegraphics[trim={0 0 0 1.15cm},clip,width = 1\textwidth]{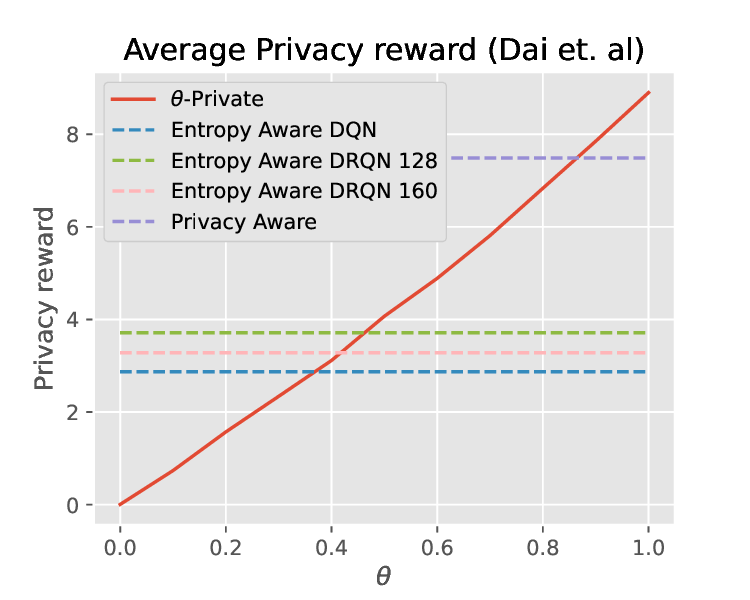}
    \caption{Heuristic privacy metric in \cite{dai2018learning}}
    \label{fig:delay}
\end{subfigure}\caption{Comparison of privacy metrics for the agents}\label{fig:res2}
\end{figure*}

The simulation setup is configured according to \cite{dai2018learning}. The mobile user buffer size is set to $b_{\text{max}}=5$ and the maximum number of newly generated tasks at each timestep is $d_{\text{max}} = 3$. It is worth mentioning that given the local processing power and the $b_{max}$ and $d_{max}$ the mobile user will have sufficient computational power to process both the task in the buffer and newly generated tasks locally  if decided.  Furthermore, it is assumed that the task size  is $M = 500 \text{ kb}$ and the mobile device operates at $F = 2 \text{ GHz}$, hence, the energy consumption for one locally processed task is $e_l = 1 \text{ J}$. The transmission speed is fixed at $r = 5000 \text{ kb/sec}$, which requires $e_1 = 0.5\text{ J}$ ($e_0 = 2 \text{ J}$) when $g=1$ ($g=0$)\cite{you2016energy}. Additionally it is assumed that the task workload density is $\eta = 500 \text{ cycles/ bit}$.  The entropies are  approximated based on the previous $W=128(160)$ state and action tuples. The length of trajectory for DRQN training is set at $N=128$. The channel state transition probability is set to $p(g_{n+1} =1|g_n = 1) = p(g_{n+1} =0|g_n = 0) =0.95$ and data processing task generation transition probability is set tot $p(d_{n+1}| d_n) = \frac{1}{1+ d_{\text{max}}}$. And finally the number of task in the buffer is determined according the action $b_{n+1} = q_n$. The delay-energy trade-off coefficient is $w_q = 0.8$. The discounting factor and the learning rate   is set to  $\alpha = 10^{-4}$, $\gamma = 0.9$. The RNN in the DRQN algorithm consists of three GRU layers with 128 hidden units and a fully connected layer on top with the total number of actions on top. The details of model parameters and architecture is given in Table~\ref{tab:sys_param}. The fully connected layer has the RELU activation function($u = max(0, a)$). Since it is practically impossible to train RNN with long sequences due to the vanishing/ exploding gradient, the input is segmented  into bundles of  16. This means that the input of the RNN unit is the most recent 16 states and actions. This technique helps to scale up and learn the same reward function in larger state space requiring longer sequences.

\begin{table*}[t]
\caption{Comparison of performances of  the trained  models and the baselines}
\small
    \centering
 \begin{tabular}{c  c ccccc}
 \toprule

 \small Agent &\small $\bar C$& \small$\bar L$(s/task) & \small $\bar E$(J/task) & H(D,T) & H(G,T) & \footnotesize privacy aware metric\\\hline
Entropy aware DQN	 & 1.663 & 0.163 & 1.499 & 3.258 & 2.019 & 2.872\\
Entropy aware DRQN(128) 	 & 1.95 & 0.411 & 1.539 & 4.481 & 3.335 & 3.713\\
Entropy aware DRQN(160) 	 & 1.878 & 0.338 & 1.54 & \textbf{4.502} & \textbf{3.338} & 3.28\\
None Private 		& \textbf{0.861} & 0.112 & \textbf{0.748} & 2.623 & 1.818 & 0.004\\
Pravacy Aware DQN (He et. al) 	 & 1.273 & \textbf{0.104} & 1.169 & 2.405 & 1.373 & \textbf{7.489}\\
\bottomrule
\end{tabular}
    \label{tab:Eval}
\end{table*}
The agents were trained for 1000 episodes and each episode lasting 1200 timesteps. As shown in Figure~\ref{fig:lc}, both Entropy Aware models used with  DQN and DRQN algorithms were able to explore the environment and converge in terms of total collected rewards in each episode. Furthermore, the DRQN agents has achieved a better total collected reward per episode which supports the use of RNNs for learning the optimal policy. As expected the model trained on trajectory length of 160 outperforms the one with trajectory length of 128 because the model observes how longer sequence of state and action patterns affect the reward function.
The performance of aforementioned agents are compared with an agent trained on the heuristic privacy metric of \cite{he2017privacy}(privacy aware) and in order further break down the entropic privacy reward it is compared against the $\theta$-private agent defined in \cite{he2017privacy}. The $\theta$-private agent acts uniformly random with probability of $\theta \in [0, 1]$ and takes the cost effective decision without any privacy concerns with probability of $1-\theta$. Intuitively, with a larger $\theta$ the agent is more private and less efficient.

\begin{figure*}[t]
\begin{subfigure}[]{.32\textwidth}
    \centering
    \includegraphics[trim={0 0 0 1.15cm},clip,width = 1\textwidth]{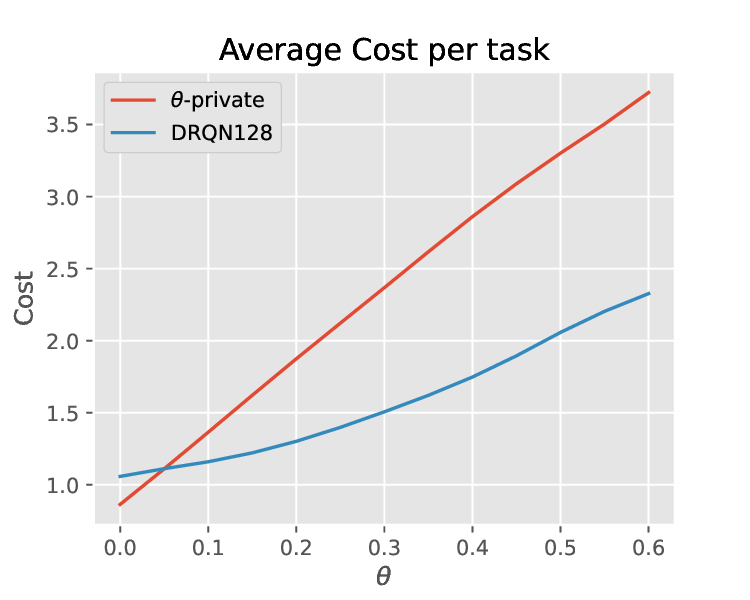}
    \caption{Cost}
    \label{fig:lambda_Hdt}
\end{subfigure}
\begin{subfigure}[]{.32\textwidth}
    \centering
    \includegraphics[trim={0 0 0 1.15cm},clip,width = 1\textwidth]{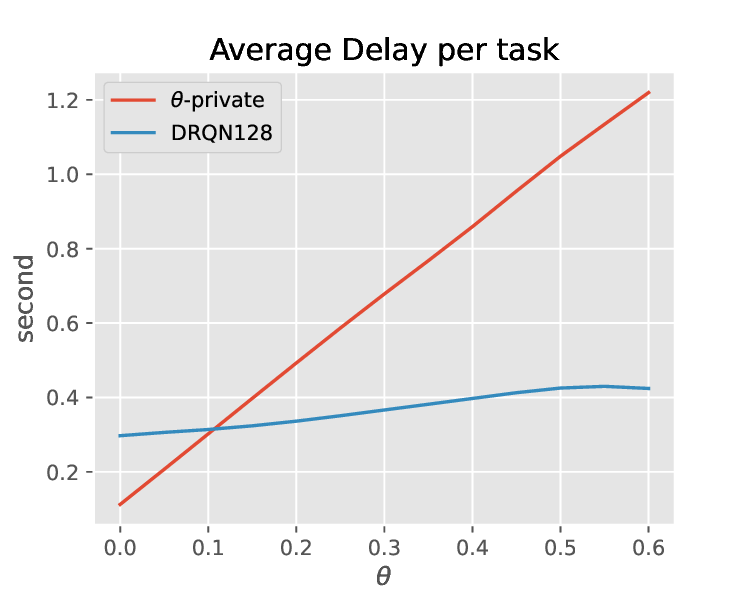}
    \caption{Delay}
    \label{fig:lambda_Hgt}
\end{subfigure}
\begin{subfigure}[]{.32\textwidth}
    \centering
    \includegraphics[trim={0 0 0 1.15cm},clip,width = 1\textwidth]{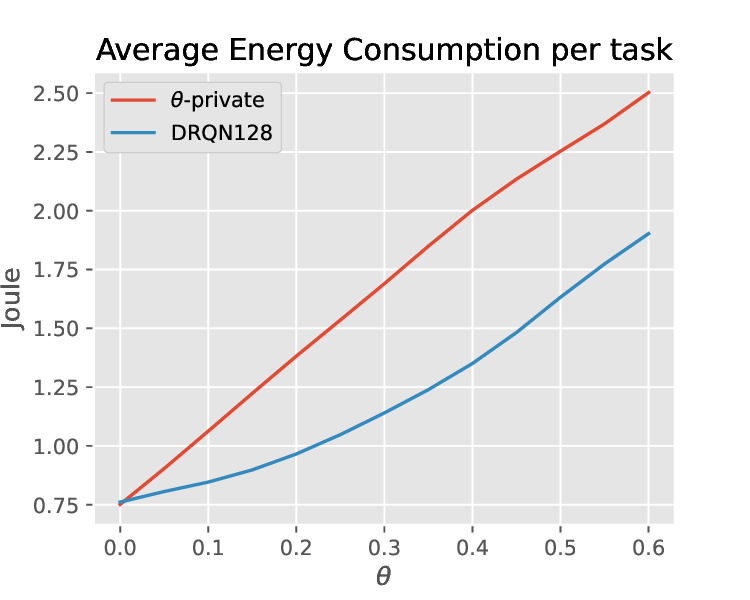}
    \caption{Energy consumption}
    \label{fig:lambda_cost}
\end{subfigure}\caption{Privacy reward weight matched to $\theta$}\label{fig:res3}
\end{figure*}
\begin{table*}[t]
\caption{Comparison of performances of  the trained DRQN128  models with different privacy reward weights}
\small
    \centering
 \begin{tabular}{c  c ccccc}
 \toprule

 \small $\lambda$ &\small $\bar C$& \small$\bar L$(s/task) & \small $\bar E$(J/task) & H(D,T) & H(G,T) & \footnotesize Privacy aware  Metric \\\hline
 2	& 1.538 & 0.601 & 0.938 & 3.649 & 2.574 & 0.536\\
5 	 & 1.596 & 0.392 & 1.204 & 4.114 & 2.98 & 1.581\\
8 	 & 1.709 & 0.299 & 1.41 & 4.318 & 3.202 & 2.364\\
10	&  1.95 & 0.411 & 1.539 & 4.481 & 3.335 & 3.713\\
16  & 2.275 & 0.422 & 1.853 & 4.675 & 3.513 & 6.474\\
20 	 & 2.454 & 0.418 & 2.036 & 4.734 & 3.598 & 8.4\\
\bottomrule
\end{tabular}
    \label{tab:Eval2}
\end{table*}

The Figure~\ref{fig:res1} compares the performance of the trained agents with the $\theta$-private agent. Both DQN and DRQN agents learn to surpass privacy aware agent in terms of energy consumption and task processing latency. As expected with a small enough $\theta$ none of the agents can compete with the $\theta$-private agent, however, shown in Figure~\ref{fig:res2}, the agent trained with the heuristic metric and the DQN agent actually does worse with regards to $H(D,T)$ and $H(G,T)$ compared to the $\theta$-private agent. This means even though the heuristic privacy reward incentivizes non cost effective action, it comes with apparent loopholes that the agent exploits towards maximising the total collected reward but the actions are not less predictable. This issue is most visible in Figure~\ref{fig:res2}(c) where the privacy aware agent does better than the $\theta$-private agent for $\theta<0.8$ while it is not doing well with regards to entropy measure at the same time. Therefore, the simulated results show that the entropy reward better encapsulates the privacy concerns in the wireless communication and is an effective tool for boosting RL agents towards acting privately in MEC networks. The performances of the agents are summarized in Table~\ref{tab:Eval} , where the  Entropy aware DRQN agents outperform the rest in the Privacy while maintaining comparable costs.
Furthermore, in the proposed reward function the privacy weight $\lambda$ is going to balance the costs (energy and delay) with the agent's privacy level. compared to $\theta$ when $\lambda \to \infty$ ($\to 0$) is equivalent of $\theta=1$(0) which corresponds to maximum privacy(zero privacy) situation. One way to evaluate the proposed model is to match $\lambda$ to $\theta$ and observe at a given privacy level which model achieves lower cost. \\

Figure~\ref{fig:res3} depicts how privacy reward weight would change the behaviour of trained agent. The agent outperforms the $\theta$-private agent for $\theta>0.04$ this means except for the  non private case, the entropy aware agent outperforms the $\theta$-private agent in terms of cost efficiency. Corollary to these arguments is that at any cost level the proposed model achieves higher privacy compared to $\theta$-private agent. Given these results it can be argued the entropy aware agent both remains cost effective with near real time processing while acting privately. Furthermore, it shows that except for the extremely high entropies   replicated with solely random decisions  for any privacy level there exists a $\lambda$ that guarantees the privacy while being cost efficient as well. $\theta$-private agent, while performing not worse in terms of the efficiency. Table~\ref{tab:Eval2} reports the average performance of Entropy aware agent (DRQN128) with different $\lambda$ values. The results show that the model is capable of adapting to the privacy preference of the user and act accordingly while maintaining the costs relatively low.

\section{Conclusion}\label{sec:conc}
In this work, the privacy issues emerged from the wireless communication in the MEC network are further investigated. A novel entropic privacy metric is introduced and shown how it contributes to the agent acting more privately in the MEC network. Moreover, a DRQN model is proposed  as a solution to the entropic privacy reward. The entropy aware DRQN agent is validated and the simulated results show the DRQN model learns the privacy reward while maintaining high levels cost efficiency. This work can be applied in the fog computing problem and help with the privacy of the IoT devices when they rely on the computation resources near the edge of the network.  Exploring the MEC network in a more general settings  such as multi user scenario, non fixed task sizes, ... as  well as considering task dropping in the action space are possible future directions of this work.

\bibliographystyle{imsart-number}
\bibliography{references.bib}

@article{nguyen2020privacy,
  title={Privacy-preserved task offloading in mobile blockchain with deep reinforcement learning},
  author={Nguyen, Dinh C and Pathirana, Pubudu N and Ding, Ming and Seneviratne, Aruna},
  journal={IEEE Transactions on Network and Service Management},
  volume={17},
  number={4},
  pages={2536--2549},
  year={2020},
  publisher={IEEE}
}

@article{qiu2019online,
  title={Online deep reinforcement learning for computation offloading in blockchain-empowered mobile edge computing},
  author={Qiu, Xiaoyu and Liu, Luobin and Chen, Wuhui and Hong, Zicong and Zheng, Zibin},
  journal={IEEE Transactions on Vehicular Technology},
  volume={68},
  number={8},
  pages={8050--8062},
  year={2019},
  publisher={IEEE}
}

@article{luo2020collaborative,
  title={Collaborative Data Scheduling for Vehicular Edge Computing via Deep Reinforcement Learning},
  author={Luo, Quyuan and Li, Changle and Luan, Tom H and Shi, Weisong},
  journal={IEEE Internet of Things Journal},
  year={2020},
  publisher={IEEE}
}

@inproceedings{he2017privacy,
  title={Privacy-aware offloading in mobile-edge computing},
  author={He, Xiaofan and Liu, Juan and Jin, Richeng and Dai, Huaiyu},
  booktitle={GLOBECOM 2017-2017 IEEE Global Communications Conference},
  pages={1--6},
  year={2017},
  organization={IEEE}
}

@article{dai2018learning,
  title={Learning-based privacy-aware offloading for healthcare IoT with energy harvesting},
  author={Min, Minghui and Wan, Xiaoyue and Xiao, Liang and Chen, Ye and Xia, Minghua and Wu, Di and Dai, Huaiyu},
  journal={IEEE Internet of Things Journal},
  volume={6},
  number={3},
  pages={4307--4316},
  year={2018},
  publisher={IEEE}
}

@article{islam2015internet,
  title={The internet of things for health care: a comprehensive survey},
  author={Islam, SM Riazul and Kwak, Daehan and Kabir, MD Humaun and Hossain, Mahmud and Kwak, Kyung-Sup},
  journal={IEEE access},
  volume={3},
  pages={678--708},
  year={2015},
  publisher={IEEE}
}

@inproceedings{liu2010tradeoff,
  title={Tradeoff between energy savings and privacy protection in computation offloading},
  author={Liu, Jibang and Kumar, Karthik and Lu, Yung-Hsiang},
  booktitle={2010 ACM/IEEE International Symposium on Low-Power Electronics and Design (ISLPED)},
  pages={213--218},
  year={2010},
  organization={IEEE}
}

@article{abbas2017mobile,
  title={Mobile edge computing: A survey},
  author={Abbas, Nasir and Zhang, Yan and Taherkordi, Amir and Skeie, Tor},
  journal={IEEE Internet of Things Journal},
  volume={5},
  number={1},
  pages={450--465},
  year={2017},
  publisher={IEEE}
}

@article{he2018deep,
  title={Deep PDS-learning for privacy-aware offloading in MEC-enabled IoT},
  author={He, Xiaofan and Jin, Richeng and Dai, Huaiyu},
  journal={IEEE Internet of Things Journal},
  volume={6},
  number={3},
  pages={4547--4555},
  year={2018},
  publisher={IEEE}
}

@article{he2019peace,
  title={Peace: Privacy-preserving and cost-efficient task offloading for mobile-edge computing},
  author={He, Xiaofan and Jin, Richeng and Dai, Huaiyu},
  journal={IEEE Transactions on Wireless Communications},
  volume={19},
  number={3},
  pages={1814--1824},
  year={2019},
  publisher={IEEE}
}

@article{kaelbling1998planning,
  title={Planning and acting in partially observable stochastic domains},
  author={Kaelbling, Leslie Pack and Littman, Michael L and Cassandra, Anthony R},
  journal={Artificial intelligence},
  volume={101},
  number={1-2},
  pages={99--134},
  year={1998},
  publisher={Elsevier}
}

@inproceedings{hausknecht2015deep,
  title={Deep recurrent q-learning for partially observable mdps},
  author={Hausknecht, Matthew and Stone, Peter},
  booktitle={2015 aaai fall symposium series},
  year={2015}
}

@article{hu2015mobile,
  title={Mobile edge computing—A key technology towards 5G},
  author={Hu, Yun Chao and Patel, Milan and Sabella, Dario and Sprecher, Nurit and Young, Valerie},
  journal={ETSI white paper},
  volume={11},
  number={11},
  pages={1--16},
  year={2015}
}

@article{mao2017survey,
  title={A survey on mobile edge computing: The communication perspective},
  author={Mao, Yuyi and You, Changsheng and Zhang, Jun and Huang, Kaibin and Letaief, Khaled B},
  journal={IEEE Communications Surveys \& Tutorials},
  volume={19},
  number={4},
  pages={2322--2358},
  year={2017},
  publisher={IEEE}
}

@article{mach2017mobile,
  title={Mobile edge computing: A survey on architecture and computation offloading},
  author={Mach, Pavel and Becvar, Zdenek},
  journal={IEEE Communications Surveys \& Tutorials},
  volume={19},
  number={3},
  pages={1628--1656},
  year={2017},
  publisher={IEEE}
}

@misc{ahmed2017mobile,
  title={Mobile edge computing: opportunities, solutions, and challenges},
  author={Ahmed, Ejaz and Rehmani, Mubashir Husain},
  year={2017},
  publisher={Elsevier}
}

@article{wu2019trust,
  title={A trust-aware task offloading framework in mobile edge computing},
  author={Wu, Dexiang and Shen, Guohua and Huang, Zhiqiu and Cao, Yan and Du, Tianbao},
  journal={IEEE Access},
  volume={7},
  pages={150105--150119},
  year={2019},
  publisher={IEEE}
}

@article{xiong2018mobile,
  title={When mobile blockchain meets edge computing},
  author={Xiong, Zehui and Zhang, Yang and Niyato, Dusit and Wang, Ping and Han, Zhu},
  journal={IEEE Communications Magazine},
  volume={56},
  number={8},
  pages={33--39},
  year={2018},
  publisher={IEEE}
}

@article{xu2019become,
  title={BeCome: Blockchain-enabled computation offloading for IoT in mobile edge computing},
  author={Xu, Xiaolong and Zhang, Xuyun and Gao, Honghao and Xue, Yuan and Qi, Lianyong and Dou, Wanchun},
  journal={IEEE Transactions on Industrial Informatics},
  volume={16},
  number={6},
  pages={4187--4195},
  year={2019},
  publisher={IEEE}
}

@inproceedings{you2016multiuser,
  title={Multiuser resource allocation for mobile-edge computation offloading},
  author={You, Changsheng and Huang, Kaibin},
  booktitle={2016 IEEE Global Communications Conference (GLOBECOM)},
  pages={1--6},
  year={2016},
  organization={IEEE}
}

@article{chen2015efficient,
  title={Efficient multi-user computation offloading for mobile-edge cloud computing},
  author={Chen, Xu and Jiao, Lei and Li, Wenzhong and Fu, Xiaoming},
  journal={IEEE/ACM Transactions on Networking},
  volume={24},
  number={5},
  pages={2795--2808},
  year={2015},
  publisher={IEEE}
}

@inproceedings{yi2015security,
  title={Security and privacy issues of fog computing: A survey},
  author={Yi, Shanhe and Qin, Zhengrui and Li, Qun},
  booktitle={International conference on wireless algorithms, systems, and applications},
  pages={685--695},
  year={2015},
  organization={Springer}
}

@book{sutton2018reinforcement,
  title={Reinforcement learning: An introduction},
  author={Sutton, Richard S and Barto, Andrew G},
  year={2018},
  publisher={MIT press}
}

@article{you2016energy,
  title={Energy-efficient resource allocation for mobile-edge computation offloading},
  author={You, Changsheng and Huang, Kaibin and Chae, Hyukjin and Kim, Byoung-Hoon},
  journal={IEEE Transactions on Wireless Communications},
  volume={16},
  number={3},
  pages={1397--1411},
  year={2016},
  publisher={IEEE}
}

@article{feng2017ave,
  title={AVE: Autonomous vehicular edge computing framework with ACO-based scheduling},
  author={Feng, Jingyun and Liu, Zhi and Wu, Celimuge and Ji, Yusheng},
  journal={IEEE Transactions on Vehicular Technology},
  volume={66},
  number={12},
  pages={10660--10675},
  year={2017},
  publisher={IEEE}
}

@article{ren2018querying,
  title={Querying in internet of things with privacy preserving: Challenges, solutions and opportunities},
  author={Ren, Hao and Li, Hongwei and Dai, Yuanshun and Yang, Kan and Lin, Xiaodong},
  journal={IEEE Network},
  volume={32},
  number={6},
  pages={144--151},
  year={2018},
  publisher={IEEE}
}

@article{alrawais2017fog,
  title={Fog computing for the internet of things: Security and privacy issues},
  author={Alrawais, Arwa and Alhothaily, Abdulrahman and Hu, Chunqiang and Cheng, Xiuzhen},
  journal={IEEE Internet Computing},
  volume={21},
  number={2},
  pages={34--42},
  year={2017},
  publisher={IEEE}
}

@article{porambage2016quest,
  title={The quest for privacy in the internet of things},
  author={Porambage, Pawani and Ylianttila, Mika and Schmitt, Corinna and Kumar, Pardeep and Gurtov, Andrei and Vasilakos, Athanasios V},
  journal={IEEE Cloud Computing},
  volume={3},
  number={2},
  pages={36--45},
  year={2016},
  publisher={IEEE}
}

@article{mnih2015human,
  title={Human-level control through deep reinforcement learning},
  author={Mnih, Volodymyr and Kavukcuoglu, Koray and Silver, David and Rusu, Andrei A and Veness, Joel and Bellemare, Marc G and Graves, Alex and Riedmiller, Martin and Fidjeland, Andreas K and Ostrovski, Georg and others},
  journal={nature},
  volume={518},
  number={7540},
  pages={529--533},
  year={2015},
  publisher={Nature Publishing Group}
}

@inproceedings{li2019learning,
  title={Learning-based pricing for privacy-preserving job offloading in mobile edge computing},
  author={Li, Lingxiang and Siew, Marie and Quek, Tony QS},
  booktitle={ICASSP 2019-2019 IEEE International Conference on Acoustics, Speech and Signal Processing (ICASSP)},
  pages={4784--4788},
  year={2019},
  organization={IEEE}
}

@inproceedings{zhao2020privacy,
  title={A Privacy-Preserving Computation Offloading Method Based on Privacy Entropy in Multi-access Edge Computation},
  author={Zhao, Xing and Peng, Jianhua and Li, Yingle and Li, Haitao},
  booktitle={2020 IEEE 6th International Conference on Computer and Communications (ICCC)},
  pages={1016--1021},
  year={2020},
  organization={IEEE}
}

@inproceedings{he2019physical,
  title={Physical-layer assisted privacy-preserving offloading in mobile-edge computing},
  author={He, Xiaofan and Jin, Richeng and Dai, Huaiyu},
  booktitle={ICC 2019-2019 IEEE International Conference on Communications (ICC)},
  pages={1--6},
  year={2019},
  organization={IEEE}
}

@article{wu2020cooperative,
  title={A cooperative computing strategy for blockchain-secured fog computing},
  author={Wu, Di and Ansari, Nirwan},
  journal={IEEE Internet of Things Journal},
  volume={7},
  number={7},
  pages={6603--6609},
  year={2020},
  publisher={IEEE}
}

@article{xiao2020reinforcement,
  title={A reinforcement learning and blockchain-based trust mechanism for edge networks},
  author={Xiao, Liang and Ding, Yuzhen and Jiang, Donghua and Huang, Jinhao and Wang, Dongming and Li, Jie and Poor, H Vincent},
  journal={IEEE Transactions on Communications},
  volume={68},
  number={9},
  pages={5460--5470},
  year={2020},
  publisher={IEEE}
}

@article{luo2020blockchain,
  title={Blockchain-enabled software-defined industrial internet of things with deep reinforcement learning},
  author={Luo, Jia and Chen, Qianbin and Yu, F Richard and Tang, Lun},
  journal={IEEE Internet of Things Journal},
  volume={7},
  number={6},
  pages={5466--5480},
  year={2020},
  publisher={IEEE}
}

@article{xiong2018cloud,
  title={Cloud/fog computing resource management and pricing for blockchain networks},
  author={Xiong, Zehui and Feng, Shaohan and Wang, Wenbo and Niyato, Dusit and Wang, Ping and Han, Zhu},
  journal={IEEE Internet of Things Journal},
  volume={6},
  number={3},
  pages={4585--4600},
  year={2018},
  publisher={IEEE}
}

\end{document}